
\input harvmac
\input epsf

\def\CZ{{\cal Z}}
\def\CS{{\cal S}}

\def\CV{{\cal V}}
\def\p{\partial}

\def\CD{{\cal D}}
\def\R{\relax{\rm I\kern-.18em R}}
\font\cmss=cmss10 \font\cmsss=cmss10 at 7pt
\def\Z{\relax\ifmmode\mathchoice
{\hbox{\cmss Z\kern-.4em Z}}{\hbox{\cmss Z\kern-.4em Z}}
{\lower.9pt\hbox{\cmsss Z\kern-.4em Z}}
{\lower1.2pt\hbox{\cmsss Z\kern-.4em Z}}\else{\cmss Z\kern-.4em Z}\fi}
\def\pl{{\it  Phys. Lett. }}
\def\prl{{\it  Phys. Rev. Lett. }}
\def\mpl{{\it Mod. Phys.   Lett. }}
\def\np{{\it Nucl. Phys. }}
\def\pr{{\it Phys.Rev. }}
\def\cmp{{\it Comm. Math. Phys. }}

\def\r{{\rm Re}}
\def\i{{\rm Im}}

\def\pt{\p_{\tau}}
\def\CM{{\cal M}}

\ifx\epsfbox\UnDeFiNeD\message{(NO epsf.tex, FIGURES WILL BE IGNORED)}
\def\figin#1{\vskip2in}
\else\message{(FIGURES WILL BE INCLUDED)}\def\figin#1{#1}\fi
\def\tfig#1{{\xdef#1{Fig.\thinspace\the\figno}}
Fig.\thinspace\the\figno \global\advance\figno by1}
\def\ifig#1#2#3{\goodbreak\midinsert\figin{\centerline{
{\epsfbox {#3.eps} }}}
\smallskip\centerline{\vbox{\baselineskip12pt
\hfil\footnotefont{\bf #1:} #2 \hfil}}
\bigskip\endinsert}
\lref\wi{E. Witten, {\it Surv. in Diff. Geom. } 1 (1991) 243.}
\lref\ackm{J. Ambj{\o}rn, L. Chekhov, C. Kristjansen and Yu. Makeenko, \np B
404 (1993) 127.}
\lref\iz{C. Itzykson and Z.-B. Zuber, {\it  Int. Journ. Mod. Phys.} A 7 (1992)
5661.}
\lref\bkdsgm{E. Br\'ezin and V. Kazakov, \pl B 236 (1990) 144; M. Douglas and
S. Shenker, \np B 335
(1990) 635;  D. Gross and A.A. Migdal, \prl 64 (1990) 127; \np B 340 (1990)
333.}
 \lref\zw{ T. Kugo and K. Suehiro, \np B337 (1990) 434; B. Zwiebach, \np B390
(1993) 33.}
 \lref\bipz{E. Br\'ezin, C. Itzykson, G. Parisi and J.-B.
Zuber, {\it Comm. Math. Phys.} 59 (1978) 35.}
\lref\mch{C. Itzykson and J.-B. Zuber, {\it J. Math. Phys.} 21, (1980) 120; M.
L. Mehta,
{\it Comm. Math. Phys.} 79 (1981) 327; S. Chadha, G. Mahoux and M.L. Mehta,
{\it J. Phys. } A:
Math. Gen. 14 (1981).}
   \lref\kdv{M. Douglas, \pl 238 B (1990) 176; T. Banks, M. Douglas, N.
Seiberg and S. Shenker, \pl 238 B (1990) 279. }
 \lref\ajm{J. Ambj{\o}rn, J. Jurkiewich and Yu. Makeenko, \pl B 251 (1990) 517}
\lref\km{ V. Kazakov and A.A. Migdal; \np B 311 (1989) 171.}
 \lref\prop{\ I. Kostov,\pl 215 B (1988) 499 ; \ D. Gross , I. Klebanov and M.
Newman, \np B 350
(1990) 621. }
\lref\pas{V. Pasquier, \np B 285 (1987) 162;  {\it J. Phys.}
35 (1987)  5707}
\lref\Inonr{I. Kostov, \pl  B 266 (1991) 317.}
\lref\adem{I. Kostov, \pl  B 297 (1992)74.}
\lref\frm{   E. Br\'ezin, V. Kazakov and Al. B. Zamolodchikov, \np  B 338
(1990) 673;
D. Gross and N. Miljkovi\chech c, \pl 238 B(1990) 217;
G. Parisi, \pl B 238 (1990) 209; P. Ginsparg
and J. Zinn-Justin, \pl B240 (1990) 333.}
 \lref\moor{G. Moore, \np B 368 (1992) 557.}
\lref\Icar{I. Kostov, ``Strings embedded in Dynkin diagrams, in the  Proc.
of the Cargese meeting {\it Random Surfaces, Quantum Gravity and Strings},
Saclay Preprint SPhT/90-133.}

\lref\Imat{I. Kostov and M. Staudacher, \np B  384 (1992) 459.}
 \lref\Iade{I. Kostov, \np B 326, (1989) 583.}
\lref\ooo{I. Kostov, \pl\   238  B (1990) 181.}
\lref\sft{ M. Kaku and K. Kikkawa, \pr D 10 (1974) 1110, 1823;
W. Siegel, \pl B151 (1985) 391; 396; {\it Introduction to String Field Theory},
World Scientific,
Singapore, 1988;   E. Witten, \np B 268 (1986) 253, A. Neveu and P. West, \pl B
168 (1986) 192.}
\lref\zinn{P. Di Francesco, P. Ginsparg and J. Zinn-Justin, 2D gravity and
 random surfaces, Preprint SPhT/93-061, to appear in {\it  Physics reports}.}
 \lref\mss{G.Moore,
N. Seiberg and M. Staudacher, \np B 362 (1991) 665.}
\lref\ms{G. Moore and N. Seiberg, {\it Int. J.  Mod. Phys.} A 7 (1992) 2601.}
 \lref\ksks{I. Kostov and M. Staudacher,  \pl B 305 (1993) 43.}
   \lref\Idis{I.K. Kostov,  \np B 376 (1992) 539.}
\lref\Iope{I. Kostov, \pl B 344 (1995) 135.}
 \lref\IM{I. Kostov and M. Staudacher, \pl B 305 (1993) 43.}
 \lref\polch{J. Polchinski \np B 362 (1991) 125.}
 \lref\dvv{M. Fukuma, H. Kawai and R. Nakayama, {\it Int. J. Mod. Phys.} A 6
(1991) 1385;
R. Dijkgraaf, H. Verlinde and E. Verlinde, \np B 348 (1991) 435.}
\lref\kmmm{S. Kharchev, A. Marshakov, A. Mironov, A. Morozov and
S. Pakuliak, \np B 404 (1993) 717.}
 \lref\kko{ V. Kazakov and I. Kostov, \np B 386 (1992) 520.}
\lref\onn{I. Kostov, \mpl A 4 (1989) 217, M. Gaudin and I. Kostov, \pl B220
(1989) 200,
I. Kostov and M. Staudacher, \np B 384 (1992), B. Eynard and J. Zinn-Justin,
\np  B 386
(1992) 459; Phys. Lett. B 302 (1993) 396.}
\Title{}
  {\vbox{\centerline
 {Feynman rules for  string
field theories}
 \vskip2pt
\centerline{
 with discrete target space
 }}}
 \bigskip\centerline{
S. Higuchi $^{a,b}$
\footnote{$^{\bullet}$}{e-mail: hig@rice.c.u-tokyo.ac.jp}
and
I. K. Kostov $^a$
\footnote{$^{\ast } $}{On leave from the Institute for Nuclear Research and
Nuclear Energy,
 72 Boulevard Tsarigradsko Chauss\'ee, 1784 Sofia, Bulgaria}
\footnote{$^{\diamond}$}{e-mail: kostov@amoco.saclay.cea.fr}}
\bigskip\centerline{{\it Service de
 Physique Th\'eorique}  \footnote{$^{\dagger}$}{
Laboratoire de la Direction
 des Sciences de la Mati\`ere du
Commissariat \`a l'Energie Atomique}{\it  de Saclay}$^{a}$ }
\centerline{{\it CE-Saclay, F-91191 Gif-sur-Yvette, France}}
\bigskip\centerline{
        {\it Department of Pure and Applied Sciences, University of Tokyo}$^b$
        \footnote{$^{\circ}$}{Present address}}
\centerline{{\it 3-8-1, Komaba, Meguro, Tokyo 153, Japan}}

\vskip .3in

\baselineskip8pt{
We  derive a minimal set of Feynman rules for the loop amplitudes in unitary
models of closed strings, whose target space is a simply laced (extended)
Dynkin diagram. The string field Feynman graphs are composed of propagators,
vertices (including tadpoles) of all topologies, and leg factors for the
macroscopic loops. A vertex of given topology  factorizes into a fusion
coefficient for the matter fields and an intersection number associated with
the corresponding punctured surface. As illustration we obtain explicit
expressions for the genus-one tadpole and the genus-zero four-loop amplitude.
 }

\bigskip
\bigskip
\bigskip
\leftline{Submitted for publication to: {\sl   Physics Letters B}}%
\rightline{ SPhT/95-056}
\Date{05/95}

\leftline{ \bf 1. Introduction}
\smallskip

One of the most important  problems in a theory of strings is  the
 construction of the corresponding  second quantized   theory, i.e.,
a field theory in the space of loops \sft .  A minimal requirement for
a string field theory is to give simple rules for the pertirbative expansion,
i.e., a prescription how to decompose  the integral over world surfaces with
different topologies into a sum of Feynman diagrams built from string
propagators and vertices.

In the last several years the simplest  noncritical string theories were
 solved using large-$N$ techniques in matrix models
(see, for example, \zinn).  A matrix model is  essentially  a system of  free
fermions
 and the closed strings  are represented there as    collective excitations of
fermions.
 A possible way to derive the  genus expansion  in  the  string theory
  is to reformulate the
 matrix model in terms of these collective fields\foot{This is not always the
best way to attack the problem.
In the standard matrix-quantum-mechanics formulation of the $C=1$ string the
fermionic formalism
 seems to be more efficient that the  Das-Jevicki collective theory, which is
well defined only on tree level.}.
 Suitable for this purpose are the ADE and \^A\^D\^E matrix
models proposed in \adem, describing     string theories   in which
the matter degrees of freedom are labeled by the nodes of a  Dynkin
 diagram $X$\Iade .
 These  models were generalized to describe both closed
 and open strings and reformulated in terms of the collective loop fields in
ref.  \Iope.
 The resulting string field theory is identical to the one obtained
with the loop gas technique in \Idis , \kko .  The world sheet of the string
represents a triangulated surface
immersed in the graph $X$.
 The diagrammatic rules  for the interactions of the  string  fields  share
some common features
 with the explicit construction of the interaction in the
critical closed string theory \zw .  The interaction is described by a
nonpolynomial action,
 with an elementary vertex for  every higher genus amplitude.   The  vertices
are  essentially
the correlation functions for
topological gravity and have the
geometrical interpretation of punctured surfaces with various
topologies localized at a single point $x$ of the target space $X$.

The aim of the present letter is to   complete  the  results of \Iope\
 where only the general form of the  closed string vertices was obtained.
 We find here  the explicit Feynman rules for a string field theory  whose
possible backgrounds are classified by  the   ADE and \^A\^D\^E  Dynkin
diagrams.
 The basic idea of our approach is
that   the higher genus
 interactions are concentrated in  the vicinity of the edge of the eigenvalue
distribution
where the singularity is always of square root one.  Near this point the
collective fields
 can be expanded in
half-integer powers which leads naturally to the KdV picture of topological
gravity.


\newsec{ Description of the target space}

  The target spaces of these models have very similar geometrical properties,
which allows to consider them simultaneously.     The  graph  representing the
target space  $X$   consists
of a set of  nodes $x$,  and a number of bonds  $<xx'>$ between nodes.
Two nodes are called adjacent ($\sim$) on $X$ if they are connected by  a
single bond.
(We do not allow more than
one bond between two nodes.)
 The graph $X$  is completely determined by  its adjacency matrix
 \eqn\conm{A_{xx'}=\cases{
1,& if $x \sim x'$    ;\cr
0,&if  not. \cr } }
  The corresponding string   theory  has a stable vacuum  only if all
the eigenvalues of $A$ are
smaller  or equal to 2. This condition restricts the choice of possible
 graphs to  Dynkin diagrams of A,D,E and \^A,\^D,\^E types.
In this case  the adjacency matrix is diagonalized as
\eqn\eIiii{ A_{xx'} =\sum_{p\in P} S_p^x \ 2\cos (\pi p)\  S_p^{x'},  }
where $P$ is the ``momentum space" dual to $X$. The  spectrum of
momenta is of the form $p={m\over h}$
 where $h$ is the Coxeter number and the integer
 $m$ are the Coxeter exponents of the corresponding Lie algebra.   The Fourier
image of a function $f_x$
 will be denoted by $f_p$
\eqn\furr{f_x= \sum_{p\in P}S_p^x  f_p , \ \ f_p =  \sum_{x\in X} S_p^x f_x.}
 The eigenvectors $S_p$ can be interpreted as the eigenstates of a quantum
particle living
 on the Dynkin diagram.
 The string background (i.e., the disc amplitude)     is proportional to   the
Perron-Frobenius eigenvector
$S_x$  labeled by the minimal momentum $p_0$\Iade
\eqn\pffp{S_x=   S^{x}_{p_0} , \ \ p_0= {\rm min}
\{ p\ | \ \ p\in P\} .}
 For the  ADE backgrounds    $p_0={1\over h}$ and for the \^ A\^ D\^ E
backgrounds $p_0=0$.
 The  ratios
 \eqn\rara{ \chi_x^{(p)} = {  S_{p}^{x}   \over S_x}}
are in correspondence with the order parameters of the matter field
and satisfy a closed
algebra \pas
\eqn\verver{\chi_{x}^{(p')} \chi_{x}^{(p'')} =\sum_{p} C^0_{ p'p''p} \
\chi_x^{(p)} .}
The symmetric structure constants  $  C^0_{ p'p''p}$    are the fusion
coefficients
 for the order parameters   and satisfy the usual relations
\eqn\mtx{ C_{p_0 p p'}^0= \delta_{p,p'}, \  \sum_p  C_{p_1p_2p}^0 C_{pp_3p_4}^0
=\sum_p C_{p_1p_3p}^0C_{pp_2p_4}^0 .}
More generally, we define the genus $g$ fusion coefficients
 \eqn\fcns{   C_{p_1\cdots p_n}^g = \sum_x    S_x^{2-2g}
\chi^{(p_1)}_x\cdots\chi^{(p_n)}_x, }
which represent  the target-space component of the the string interaction
vertices \Iade .

\newsec{ A large $N$ matrix model  for the loop fields}

 The    string  theory  with target space $X$   will be constructed as
 the  $1/N$ expansion of the  $N\times N$
  matrix    model  defined in  ref. \adem\
    whose entities are  the hermitian matrices ${\bf H}_x $ associated
with the sites of the lattice, and the  complex   matrices
${\bf C}_{xx'}={\bf C}_{x'x} ^{\dag}$ associated with the links $<xx'>$.
The partition function of the model reads
\eqn\mpatr{\CZ[V]= \int \prod_x D{\bf H}_x\prod_x e^{-\tr  V_x ({\bf H}_x)}
\prod_{<x x'>} D{\bf C}_{xx'}
\  e^{\tr {\bf H}_x  {\bf C}_{xx'}{\bf C}_{x'x} }.}

The  operator  creating a macroscopic loop  at the point  $x$ is the resolvent
of
 the matrix field ${\bf H}_x$
\eqn\loopp{   W_x(z)= \tr \  {1\over {z-\bf H}_x}.}
The $n$-loop amplitude is obtained by differentiating with respect to the
source
 \eqn\nlp{
  \langle \prod _{s=1}^n    W_{x_s}(z_s)\rangle
 =\Bigg(     \prod_{s=1}^n \int {d\lambda _s \over\lambda_s-z_s}
{\delta\over \delta V_{x_s}(\lambda_s)}\ \ \ln\CZ[V]\Bigg)_{V=V_0 }.}
  where  $V_0$ is a  polynomial  potential that determines the vacuum state of
the theory.
 Since we restrict ourselves to the unitary models, a polynomial of third
degree $V_0(z)= g_1 z +g_2z^2 +g_3z^3 $ is sufficient.

   After integrating  with  respect to the the ${\bf C}$-fields, the partition
function
 \mpatr\ can be defined in terms of the (shifted) eigenvalues
$\lambda_{ix}, i=1,2,\ldots,N,$    of the    field  ${\bf H}_x $  as \adem
    \eqn\mtria{
\CZ[ V ] =     \int \prod _{i,x}      d\lambda_{ ix}  e^{ -V_x ( \lambda_{ix})
} \  \prod_{x,x'}
{\prod_{i\ne j} (\lambda_{ix }-\lambda_{j x'})^{\delta_{x,x'} }\over
   \prod_{i,j}  |\lambda_{ix} +\lambda_{jx'}|^{\half A_{xx'}}}.  }
 (The potential will change as well, but  we will keep the same letter for it.)

Using the Cauchy identity, we formulate this model as a system of free
fermions.
This is the way to look for a nonperturbative solution. Our aim  is to extract
the
 perturbative  piece of the partition function. For this purpose it is most
convenient
 to introduce collective field  $\psi$ and the corresponding lagrange
multiplier field $v$
describing the fluctuations of the system above the
large $N$ saddle point.
  This is done by   inserting the
identity
\eqn\jacc{1= \int \CD\psi\CD v \exp \Bigg(
\sum_x\int d\lambda \  v_x(\lambda)\Big[ {d \psi_x(\lambda) \over d \lambda }
-\sum _{i=1}^N \delta(\lambda -\lambda_{ix})\Big]\Bigg)}
in the r.h.s. of \mtria, and integrating  over the
    $\lambda$'s.  For each  $x$  the integration with respect to the
$\lambda_{ix}, i=1,\ldots,N,$ yields the  ``pure gravity" partition function
\eqn\onmm{e^{ \CF[v]}= \int  \prod_{i=1}^N   d\lambda_{i } e^{-  v(\lambda_i)}
\prod_{i<j}(\lambda_{i }-\lambda_{j })^2}
with potential $v=   v_x$.
 After that  the original partition  function \mtria\ can be written as
  a functional integral
\eqn\ppprt{\CZ[V] =\int   \CD  \psi  \CD v \
 e^{  - \CS  [\psi ,v,   V] } ,}
 \eqn\acctt{\eqalign{  \CS  [\psi ,v,   V]   &=
    \half \sum_{x,x'}\int d\lambda  d\lambda'\  A_{xx'}
{d \psi_x(\lambda)\over d\lambda}
\ln | \lambda+ \lambda'| \ {d \psi_x'(\lambda')\over d\lambda'}
    \cr
&+ \sum_x \int d\lambda {d \psi_x(\lambda)\over d\lambda}
     \Big( V_x(\lambda) -v_x(\lambda)\Big)  -\sum_x \CF   [v_x].
\cr}}
The zero mode in the $\psi$-integration is eliminated by imposing a Dirichlet
boundary condition $\psi_x(\infty)=0$.

The string coupling constant $\kappa\sim 1/N$ is contained in the genus
expansion of
 the effective potential
$\CF[v_x]= \CF_0[v_x] +\CF_1[v_x]+\cdots $.

\newsec{Saddle point}

   The functional measure in  \ppprt\ is a nonrestricted homogeneous measure
  and the string propagator and vertices  are obtained by expanding the
effective
action \acctt\ around the
 mean field $\psi_{x}^{c},    v_x^{c}$ determined by the large $N$ saddle point
equations.
The  genus-zero expectation value   of  the resolvent
\loopp \ is related to   the classical spectral density   $\rho_x^c(\lambda) =
\p_{\lambda} \psi^c _{x}=  \delta\CF_0/\delta v_x(\lambda)[v_x^c] $ by
\eqn\rezw{\langle W_x(z)\rangle_0 =\int d\lambda {  \p_{\lambda} \psi^c _{x}
\over z-\lambda}. }
Along the real axis
\eqn\rezwz{
\langle W_x(\lambda)\rangle_0 = \half\p _{\lambda} v_x^c -i\pi    \p \psi^c
_{x}  }
 and  the saddle point equations  can be written, in the momentum space,  as
\eqn\dspe{   \ 2\r \langle W_p (\lambda)\rangle_0  +2\cos \pi p\
\langle W_{p}(-\lambda)\rangle_0
 = \delta_{p,p_0} V_0(\lambda) .}

The  density    $\rho^c = \rho^c _{p_0}  $ has a compact support $[a,b]$ with
$b<0$.
We are interested  in the scaling regime, which is  achieved in the limit $a/b
\to \infty .$
To render the equations simpler, we will rescale   $ \lambda \to |b|^{-1}
\lambda $; then the
support of   $\rho^c$     becomes the  semi-infinite interval $[-\infty, -1]$.
The solutions of the saddle point  equation  \dspe\ behave at infinity as
$z^{\beta}$, where
 $\cos \pi\beta =-
\cos \pi p_0$. The two  branches
 \eqn\eqq{{\beta}=1\pm p_0   }
 correspond to the dense $(-)$ and dilute $(+)$  critical regimes of the model.
Up to an arbitrary normalization the solution in the scaling limit reads
 \eqn\pphhi{ \langle W_x (z) \rangle_0 =
- {  S_x \over \kappa} \ { (z +\sqrt{z^2-1})^{\beta} +(z -\sqrt{z
^2-1})^{\beta} \over
\  2\pi  {\beta}  |\sin \pi {\beta} | }}
where
  $\kappa  \sim   L^{1+{\beta}} / N $ is the renormalized string
coupling constant.
   A  string theory with
 this background  can be viewed as a   theory of 2d quantum gravity with the
central
charge of the matter field
\eqn\bab{ C= 1-6 {({\beta} -1)^2\over {\beta}} .}
 For details see \Idis \ and \Imat.

\newsec{Gaussian fluctuations and analytic fields}
In order to derive a set of Feynman rules one has    to define the Hilbert
space of one-string states,
 chose a complete orthonormalized  set of eigenstates of the  quadratic action,
and finally express
the interactions in terms of the mode expansion of the string field.

It is consistent with the  perturbative expansion to assume that the
fluctuating fields are again
  supported by a semi-infinite  interval, but its right end can be displaced
with respect to its
saddle point value $-1$    due to the fluctuations. Therefore
  it is  possible to represent the collective fields in terms of the analytic
functions
$\Psi_x(z)$ and $\Phi_x(z)$ defined on the $z$-plane cut along  the negative
real axis and
such that
 \eqn\anfs{   \psi _x(\lambda) =  {1\over \pi}  \i  \Psi_x( \lambda ) , \ \
v_x(\lambda)=
 2\r \Phi_x( \lambda ), \ \
\ \lambda <0}
determined up to an entire function of $z$.
 We will consider the      discontinuity
\eqn\angf{  \phi _x(\lambda) =  {1\over \pi}  \i  \Phi_x( \lambda  )}
  of the analytic  function  $ \Phi_x(z)$ as  independent field variable,
instead of $v_x$.
  The quantum field $\Psi$ is related to the
loop operator  \loopp\  by
 \eqn\LPS{W_x(z) = {\p \over \p z} \Psi_x(z)}
and  can be interpreted geometrically as the operator creating a loop without
marked point on the world sheet.
The   macroscopic loop correlator  \nlp\   is equal to the derivative $\p^n/ \p
z_1 \cdots \p z_n
$ of the connected correlator of $ \langle \Psi(z_1) \cdots \Psi(z_n)\rangle$.
In the following by loop correlators
we shall understand the correlators of the $\Psi$-field.
Since we are interested only in the scaling limit, it is consistent to identify
all analytic
 fields that differ by an entire function.

 The gaussian fluctuations of the collective fields are those that do not shift
the edge of
 the eigenvalue interval; the  fluctuations    displacing  of the edge   are
described by
 nongaussian terms in the effective action that represent  the   $n$-string
interactions.
 Therefore the leading (genus-zero) term of the effective action
\acctt\ can be split into   gaussian and interacting parts
\eqn\ssss{  \CS_0 =\CS^{\rm free}  + \sum _{n\ge 3} \CS _{0,n}  }
 where  $ \CS _{0,n}$ is the genus-zero  $n$-string  interaction.

 Taking the  genus-zero contribution to the mean field free energy  \onmm\
\eqn\flnn{\CF_0[v_x]= {1\over \pi} \int d\lambda
(\r \Phi_x - v_x   )  \i { d \Phi_x \over d \lambda} =
-{1\over \pi} \int d\lambda\r \Phi_x
   \i { d \Phi_x \over d \lambda} }
  we find for  the gaussian  action
\eqn\ccii{\eqalign{\CS^{\rm free} &= {1\over \pi} \sum_x
\int _{-\infty} ^{-1} d\lambda  \big[ \r \Phi_x (\lambda) {d\over d\lambda}
\i \Phi_x (\lambda)
  - 2\r \Phi_x(\lambda)  {d\over d\lambda}  \i \Psi_x  (\lambda)
  \cr
&+\sum_{x'} A_{xx'} \Psi_{x'} (-\lambda) {d\over d\lambda}
\i \Psi_x(\lambda)
+V_x(\lambda) {d\over d\lambda} \i \Psi_x(\lambda) \big] . \cr}}
 In writing \ccii\  we used  that $ \i\Psi_x (-\lambda)=0, \ \lambda <0$.

We define the  Hilbert space $\CH$ of one-string states as the space of real
functions
$\phi_x(\lambda)$ defined on the interval $-\infty <\lambda <-1$, with scalar
product\foot{The choice of the scalar product is a matter of convenience and
does
 not affect the final results. It is not conserved  by the linear
transformations
relating  different functional realizations  of the  one-string configuration
space \mss, \ms .}
 \eqn\sclp{ \langle \phi |\psi\rangle = \sum _{x\in X} \int_{-\infty}^{-1}
{d\lambda\over \sqrt{\lambda^2-1}} \ \psi_x(\lambda)\
\phi_x(\lambda).}

  For the purpose of diagonalizing the quadratic action it is very useful to
introduce
    the map
     \eqn\prms{z(\tau) =z(-\tau) =  \cosh \tau .  }
    transforming  the space of
meromorphic functions  in  the   $z$-plane  cut along the interval
 $[-\infty , -1]$ into  the space of entire even analytic functions of $\tau$.
The  cut $z$-plane is parametrized by  the semi-infinite strip $\{ \r \tau \ge
0, -\pi \le \i \tau  \le \pi \}$
 so that the two sides of the cut are parameterized by the boundaries $\{ \tau
\pm i\pi ,
 \tau >0 \} $ of the strip
\eqn\prpi{ \lambda = \cosh \ (\tau \pm i\pi)=-\cosh \ \tau, \ \ \tau \ge 0 . }
Due to the symmetry of the map \prms\ the contour integrals around the cut
transforms into integrals
along the shifted real axis $\tau -i\pi, \ -\infty <\tau <\infty$.
In the following we will keep the same letters for the fields considered as
functions of $\tau$ and denote
$ \Phi(\tau) \equiv \Phi(z( \tau)) $.
 The  disc amplitude \pphhi\   as a function of   $\tau$-variable is
 \eqn\ccoc{ \langle W_x(\tau)\rangle _0 =  {S_x \over  \kappa}\ \
{\cosh \beta \tau \   \over 2\pi \beta  |\sin \pi  \beta |   }= {\p  \Phi_x
^c(\tau)\over \p\cos \tau} .}

It is quite evident that the   plane waves
\eqn\fint{\langle x, \tau |p, E \rangle = S_p^x  e^{iE\tau}}
 form a
 complete set of (delta-function) normalized wave functions diagonalizing the
quadratic action \ccii .
 The Fourier components of the fields $\phi$ and $\psi$ are related to these of
$\Phi$ and $\Psi$ by
$\phi(p,E)= {1\over\pi}\sinh (\pi E ) \Phi(p,E), \psi(p,E)= {1\over\pi}\sinh
(\pi E ) \Psi(p,E)$.
The action  \ccii \ reads, in terms of these fields,
\eqn\elisav{\eqalign{ \CS^{\rm free}[\psi,  \phi]&=\sum _{p\in P}  \int
_{0}^{\infty}{dE\over 2\pi}
   \Big([\phi(p,E)- 2 \psi (p,E)    ]\ { \pi E\cosh\pi E  \over  \sinh \pi E}
 \ \phi(p,E)\cr
&+ \psi (p,E)  { \pi E\cos\pi p   \over  \sinh \pi E}  \psi (p,E)-  E  V(p,E)
\psi(p,E)     \Big) \cr}}
 By inverting the quadratic form in \elisav\ we find the
  propagators  in the $(E, p)$ space
 \eqn\prpp{\eqalign{ G_{\psi\psi}(E,p)&= G_{\phi\psi}(E,p)=  G(E,p), \cr
   G_{\phi\phi}(E,p) &=    G(E,p){\cos\pi p\over \cosh \pi E} = (G(E,p)-
G(E,\half) )\cr}}
where
\eqn\bleve{G (E,p)= { \sinh \pi E \over\pi  E  } {1\over   \cosh \pi E - \cos
\pi p } =  {2\over\pi^2}
  \sum _{n=-\infty}^{\infty}
{1 \over E^{2}+(p+2n)^{2}}.}

The three propagators have the following diagrammatic meaning:  $G_{\phi\psi} $
 ($ G_{\phi\phi}$)
 is  associated with  the external (internal) lines of a Feynman diagram,  and
 $G_{\psi\psi}$ is
 the genus-zero loop-loop correlator \Idis
\eqn\llco{\langle \Psi_p(z)\Psi_p(z')\rangle_0 =   \int _0^{\infty}
dE
{E\over \sinh\pi E} {\sin E\tau \ \sin E\tau '\over   \cosh \pi E - \cos \pi p
}}

\newsec{Interactions}

The genus $g$, $n$-string interactions, $2g+n-2>0$,  are determined by the
genus expansion of the
 one-matrix free energy $\CF[v]$ defined by \onmm .
 In the scaling limit $\CF[v]$   is the generating functional for
 the correlation functions
\eqn\crff{ \{ k_1\cdots k_n\}_g =\langle \sigma_{k_1} \cdots
\sigma_{k_n}\rangle_g }
of the scaling operators $\sigma_k, k=1,2,\ldots,$ in topological gravity
\ref\wwi{E. Witten,
 \cmp 117 (1988)) 411; 118 (1988) 411; R. Dijkgraaf and E. Witten,
\np B 342 (1990) 486 .}.
  The correlation functions \crff\  are the intersection numbers on the moduli
space $\CM_{g,n}$ of algebraic curves  of genus $g$ with $n$ marked points \wi
{}.
Each operator $\sigma_k$ represents a $2k$-form in the moduli space $\CM_{g,n}$
and
 the correlation function \crff\ is nonzero only if the product is a volume
form, i.e.,
if $2(k_1+\cdots +k_n)=  {\rm dim} \CM_{g,n}= 2(3g-3+n)$. The intersection
numbers can be
obtained from a system of reccurence relations equivalent to the loop equations
\ref\dvv{ R. Dijkgraaf, E. Verlinde and H. Verlinde, \np B348 (1991) 435;
M. Fukuma, H. Kawai and R. Nakayama, {\it Int. J. Mod. Phys.} A6 (1991) 1385}.
In particular, the genus-zero the intersection numbers coincide with the
multinomial
coefficients
\eqn\ssts{ \{  k_1 \cdots k_n \} _0 =   {(k_1+\cdots+k_n)! \over k_1!\cdots
k_n!} , \ \ k_1+\cdots+k_n= n-3.}

For generic potential  $v=2\r \Phi$, the deformation parameters $t_{0},
t_{1},\ldots,$ are
 proportional to the coefficients in the
 expansion of the analytic function $\Phi(z) $ in the half-integer powers of
$z-z_0$ for some $z_0$.
Sometimes these parameters are named KdV coordinates of the potential $v$.
The genus  $g$ term in the perturbative  expansion    of $\CF $ reads
 \eqn\pssp{\CF_{g}[\Phi]  =\kappa_0  ^{2g-2}
\sum_{n\ge 0} {1\over n!} \sum_{k_1,\ldots,k_n \ge 0} \{  k_1 \cdots k_n \} _g\
t_{k_1}\cdots   t_{k_n}, }
   where $\kappa_0$ is the string coupling and the sum is restricted to $
k_1+\cdots+k_n= 3g-3+n$.

  There is certain freedom in the choice  of the  coupling constants     and we
will use it
 in order to get simpler formulas.    We choose $z_0=-1$  and define   the
deformation
parameters   $t_{kx} =t_k(\phi_x) $
 by\foot{As  mentioned above, we are
considering $\phi = {1\over \pi} \i\Phi$ as independent variable.}
 \eqn\depr{ \delta_{k,1} - t_{kx} = {\langle k|  \phi_x\rangle\over \langle 1|
\phi^c_x\rangle}}
  where the  linear functionals   $ \langle k|  , k=0,1,\ldots,$   are
determined by the generating function
\eqn\bppyy{\sum_{k=0}^{\infty} {u^k\over k!} \langle k|  \phi\rangle =
\sqrt{2} \oint {dz\over 2\pi i} {d  \Phi/ dz  \over  \sqrt{ z+1 -2u}}}
and $\phi_x^c$ is the mean field corresponding to $v_x^c$ and $\psi_x^c$.
With this definition the first two KdV coordinates of the string background
vanish,
$t_0(\phi^c_x) =t_1(\phi^c_x)=0$, and the sum in \pssp\ contains only  finite
number of terms.
  The topological-gravity  coupling constant corresponding to the normalization
\depr\ is $\kappa_x= - \langle 1| \phi_x^c\rangle ^{-1}$.  Note that its value
generically depends on the point $x$ in the target space. This is natural,
since the mean field is $x$-dependent.  The   genus $g$, $n$-string interaction
is a linear functional in the tensor product $\CH^{\otimes n}$
\eqn\stin{\CS_{g,n}( \phi)=    \langle \CV_{g,n}|  \phi \rangle \cdots  |
\phi\rangle .}
Expanding  $\CF_{g}[\phi_x] $  around the saddle point $\phi^c_x$ we find the
vertex $\CV_{g,n}$ in an operator form
\eqn\psjsp{\langle  \CV_{g,n}| ={1\over n!}
\sum \langle 1| \phi_x^c\rangle^ {2-2g-n}
     \{ k_1 \cdots k_{n+m}    \}_g\ {t^c_{k_{n+1}}\cdots  t^c_{k_{n+m}}\over
m!}\
 \langle k_1, x|  \cdots  \langle k_n, x| ,}
  where the sum goes over $x\in X,  m=0,1,\ldots; k_1,\ldots,k_{n+m}\ge 0;
k_1+\cdots +k_{n+m} = 3g-3+n$, and $\langle k, x|\phi\rangle \equiv  \langle
k|\phi_x\rangle$.
In order to find the  functional representation of the vertex $\CV_{g,n}$ we
have to
 calculate  the KdV coordinates
  of  the plane waves \fint .
 In the $\tau$-parametrization, the generating function   \bppyy\  reads
\eqn\genn{  \sum _{n\ge 0}   {u^n \over n!} \langle n |  \phi\rangle = \sqrt{2}
\int_0^{\infty}d\tau
    {\pt  \cos \pi \pt   \phi(\tau) \over   \sqrt{\cosh\tau + 1-2u}}. }
 For  $\phi (\tau) =  \sin E\tau$  the r.h.s. of  \genn\  is a double
series in $u$ and $E$
\eqn\legg{  {\sqrt{2} \over \pi} \int_0^{\infty} d\tau{ \cosh\pi E \cos
E\tau\over \sqrt{\cosh\tau + 1-2u} }=
 P_{-\half +iE}(1-2u)=    \sum _{k=0}^{\infty}
 { u^k \over k!   }\Pi_k(   iE)  ,  }
where
\eqn\pii{ \Pi_0(a)=1; \ \ \Pi_k(a)= {1\over k!} \prod_{j=0}^{k-1}[(j+\half  )^2
-a^2], \ \ k=1,2,\ldots}
Therefore the KdV coordinates of a plane wave are
\eqn\pipi{ \langle k | E\rangle=   \ \pi  E  \ \Pi_k(iE) , \ k=0, 1,2,\ldots.
}

 The KdV coordinates $t^c_k$ of the string background can be readily found
using the fact that
the derivative $d\Phi^c(z) /dz$, eq. \ccoc,  is   a plane wave with $E=i\beta$,
and the
  identity  $ \langle  k | d\phi / d z  \rangle = \half \  \langle
k+1|\phi\rangle$. We get
 \eqn\teka{   \langle k|\phi ^c_x\rangle =  -  {S_x\over\kappa}\ (\delta _{k,1}
-t^c_k)}
with
\eqn\bertr{   t^c_0 = t^c_1 =0, \ \
   t^c_k =  -  \Pi_{k-1} ( \beta ), \ k=2,3,\ldots   }

 Combining \psjsp , \pipi\ and \bertr\ we find the explicit form of the vertex
$ \CV_{g,n}$ in the $(p,E)$  space
\eqn\vpso{\eqalign{ \langle  \CV_{g,n}|p_1, E_1\rangle \cdots  |p_n, E_n\rangle
&={ \kappa ^{2g-2+n}\over n!}
\sum _{ p_1,\ldots,p_n}    C_{p_1\cdots p_n}^g  \sum_{m\ge 0}    \cr
\sum_{k_1.. . k_{n+m}} \{ k_1\cdots  k_{n+m} \}_g
 &{t^c_{k_{n+1}}\cdots  t^c_{k_{n+m}}\over m!}\ \langle k_1| E_1\rangle  \cdots
 \langle k_n| E_n\rangle ,\cr}}
 where the sum goes over $  m=0,1,\ldots; k_1,\ldots,k_{n+m}\ge 0; k_1+\cdots
+k_{n+m} = 3g-3+n$.

The   vertex \vpso\ is represented   in the $\tau$-space  by a distribution
supported by the point $\tau =0$.
 Indeed, eq. \pipi\ means that the linear functional $\langle k| $ acts in the
space of  odd
functions  $\phi(\tau)= -\phi(-\tau)$   smooth   at
$\tau =0$ ,  as
 \eqn\pipi{  \langle k|\phi\rangle  = \pi   (\Pi_k (\pt) \ \pt \phi(\tau)
)_{\tau =0}. }
Therefore  the  scattering of string states occurs only along the edge of the
half-space $(x,\tau)$.
 This renders
the discrete-space formulation of the string theory simpler than the continuous
one
where the interaction
 only falls exponentially at $\infty$.

\newsec{Feynman rules in the KdV representation}

  The most efficient  way of calculating Feynman diagrams is to represent all
entities
by their KdV coordinates.
The KdV representation of the internal propagator is   given by
 the symmetric matrix
\eqn\intp{\eqalign{G_{kk'}(p)= &\langle k|\langle k'| G_{\phi\phi}
(p)\rangle\cr
 = &\pi^2 \Bigg(\Pi_k( {\p_\tau}) \ \Pi_{k'}({\p_{\tau'}})  {\p_\tau}
 {\p_{\tau'}}\ \Big(G_{\phi\phi} (\tau,\tau'; p)\Big)\Bigg)_{\tau=\tau'=0}
.\cr}}

 The
 coordinate representation of the  propagator \bleve\  is
 \eqn\ppiu{\eqalign{
G(\tau,\tau';p)&=
\int _0^{\infty} {4 \over\pi}
\sin E\tau\ \sin E\tau'\  G(E,p)dE\cr
 &= G(\tau-\tau'; p) - G(\tau+\tau'; p)  \cr}}
   where
 \eqn\expl{
{\p G(\tau; p ) \over \p\tau}= -{1\over \pi^2} \sum_{n\in\Z} e^{-|(p+2n)\tau|}=
 -{1\over \pi^2} {\cosh (1-p)\tau\over \sinh\tau},\ \  0\le p<2.}
The
internal  propagator $G_{\phi\phi}(\tau,\tau';p)$  is regular at the origin
  \eqn\teri{G_{\phi\phi}(\tau,\tau';p)
=  {1\over \pi^2} \Big(p-{1\over 2}\Big)\Big(p-{3\over 2}\Big)
  \tau\tau ' \Big[ 1+{1\over 12} \Big(p^2-2p-{3\over 4}\Big)
(\tau^2+\tau'^2)+\cdots\Big]}
and from its Taylor expansion   one extracts the matrix elements \intp
      \eqn\expl{\eqalign{ G_{00}(p)=& -\Pi _1(1-p), \ \ G_{01}(p)=G_{10}(p)=
-\Pi _2(1-p) ,\cr G_{11}(p)=&-2\Pi_2(1-p)-2\Pi_3(1-p), \ \ \ldots\cr}}

  The external line factors in the corresponding Feynman diagrams are given by
   \eqn\exll{\eqalign{\langle k , p| G_{\phi\psi}|\tau , p \rangle  &=
  {\pi   \over \sin \pi\pt}
\int _0^{\infty} {dE \over \pi} \sin E\tau \
 E   \ \Pi_k(iE)  G(E,p)    \cr
&=       \Pi_k({\p_\tau} )  {\sinh(1-p)\tau \over\sin \pi p\sinh\tau}\cr}.}

 The Feynman rules are summarized in \tfig\feynmanrule.
It is understood that the dressed vertices are composed by adding tadpoles to
the bare vertices.

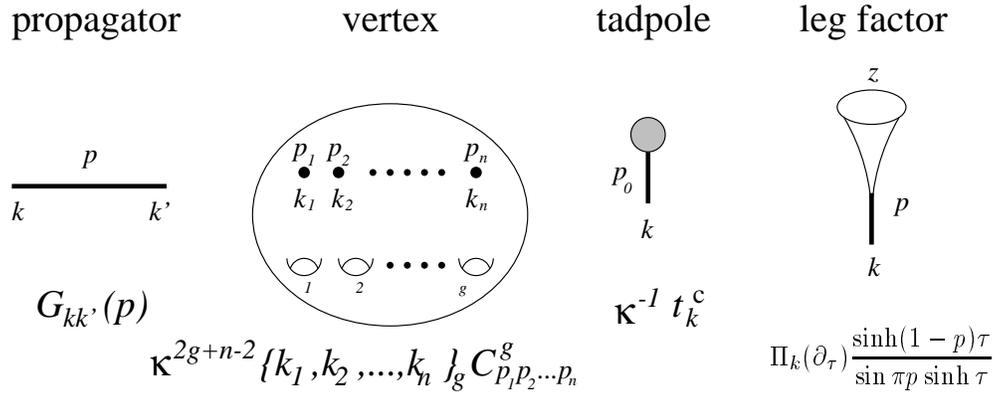
\ifig\feynmanrule{The Feynman rules for the string field theory.}{feynman}

As an illustration we present  some simple examples.

\noindent $(i)$  Let us first check that the Feynman rules  reproduce the loop
correlation functions
in the topological gravity  when $X=A_1, t^c_k =0, k\ge 0 $.
The   momentum space contains  a single momentum $p_0=1/2$, hence
the propagator $G_{\phi\phi}$ is identically zero and  the leg factors are
\eqn\TGg{ \Pi_k (\pt){\sinh (\tau/2)\over \sinh\tau} = \Big(-2{d\over
dz}\Big)^k{1\over\sqrt{2(z+1)}}.}
  In the spherical limit we get, using the explicit form of the intersection
coefficients \ssts, the
compact  formula found originally in  \ajm
 \eqn\Ttp{\eqalign{ \langle \prod_{a=1}^n \Psi(z_a)\rangle _0&= \sum
_{k_1+\cdots +k_n=n-3} {(k_1+\cdots +k_n)!\over k_1!\cdots k_n!}
\prod _{a=1}^n   \Big(-2{d\over dz_a}\Big)^{k_a} {1\over\sqrt{2(z_a+1)}}\cr
&=\Bigg( {d^{n-3}\over du^{n-3}} \prod _{a=1}^n
{1\over\sqrt{2(z_a+1-2u)}}\Bigg)_{u=0}.\cr}}

\noindent $(ii)$  The   three-loop genus-zero amplitude  is just a product of
a vertex  $\{000\}_0 C_{p_1p_2p_3}^0$ and
of three leg factors (\tfig\threeloopgenuszero)
\eqn\thrt{  \langle \prod _{s=1}^3    \Psi_{p_s} (z_s) \rangle_0= \kappa
C_{p_1p_2p_3}^0 \prod _{s=1}^3
   {\sinh (1-p_s)\tau_s \over \sin \pi p_s \sinh \tau_s}.}
One can check \ksks\ that for  $\beta = 1$ this amplitude coincides, after
being transformed to
 the $x$-space, with the
three-loop amplitude   in the   string theory with continuous target space,
calculated in \ms .

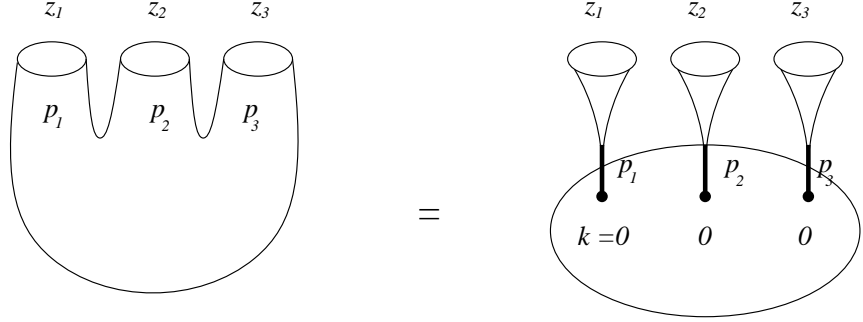
\ifig\threeloopgenuszero{three-loop genus-zero amplitude.}{3loopgenus0}

 \noindent $(iii)$  The genus-one tadpole (\tfig\genusonetadpole).
 \eqn\gotp{\eqalign{  \langle   \Psi_{p} (z) \rangle_1 &=  \kappa
\Big[  C^1_{p} \{ 1 \}_1  \Big({1\over 4}- { \p^2  \over \p  \tau^2 }\Big)
+   C^1_{p} \{ 02\}_1 t_2^c \cr
&+  {1\over 2}  \sum_{p'} C^0_{p'p'p} \{000\}_0
 G_{00}(p')  \Big]  {\sinh (1-p)\tau \over \sin \pi p \sinh \tau} \cr
  & = \kappa \delta_{p,p_0}
\sum _{p'\in P}\Big[{1\over 24}   \Big(\beta^2 - {\p^2 \over \p \tau^2}\Big)
+ {1\over 2}\Big(p'-{1\over 2}\Big)\Big(p'-{3\over 2}\Big) \Big]
{\sinh (1-p_0)\tau \over \sin\pi p_0 \sinh \tau}. \cr}}
 This expression is in accord  with the continuum limit of the genus-one loop
amplitude for the $O(n)$ model on a random lattice, $n=2\cos \pi p_0$, obtained
recently by B. Eynard and C. Kristjansen
\ref\BC{ B. Eynard and C. Kristjansen, preprint SPhT-068} (in this case  the
sum
contains  only one term, $p'=p_0$).

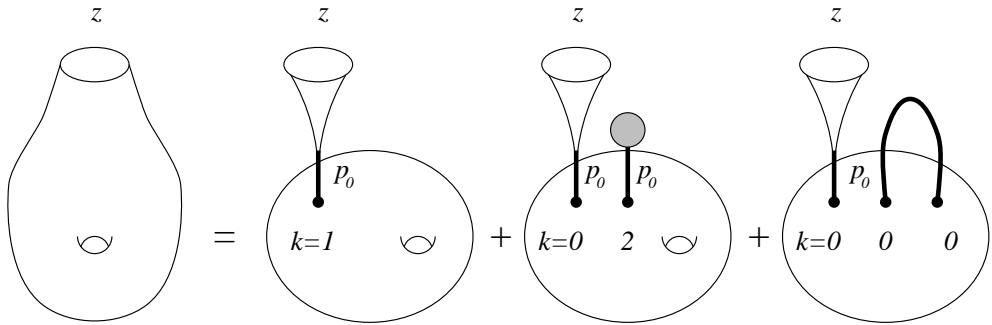
\ifig\genusonetadpole{Genus-one tadpole.}{genus1tadpole}

\noindent $(iv)$ The four-loop genus-zero amplitude
 (\tfig\fourloopgenuszero).
\eqn\ivan{\eqalign{ &\langle \prod _{s=1}^4    \Psi_{p_s}(z_s)\rangle_0=
\kappa^2  \Bigg[\Big(   \beta ^2-{1\over 4}  +  \sum_{s=1}^4
\Big({1\over 4} - {\p^2\over\p\tau_s^2}\Big)  \Big)C^0_{p_1p_2p_3p_4} \cr &
+ \sum_p  \Big( C^0_{p_1p_2p}C^0_{p p_3p_4} +C^0_{p_1p_3p}C^0_{p
p_2p_4}+C^0_{p_1p_4p}C^0_{p p_2p_3}\Big)
  \Big(p-{1\over 2}\Big)\Big(p-{3\over 2}\Big)  \Bigg] \cr &
\prod_{s=1}^4  {\sinh (1-p_s)\tau_s \over \sin \pi p_s \sinh \tau_s}.\cr}}

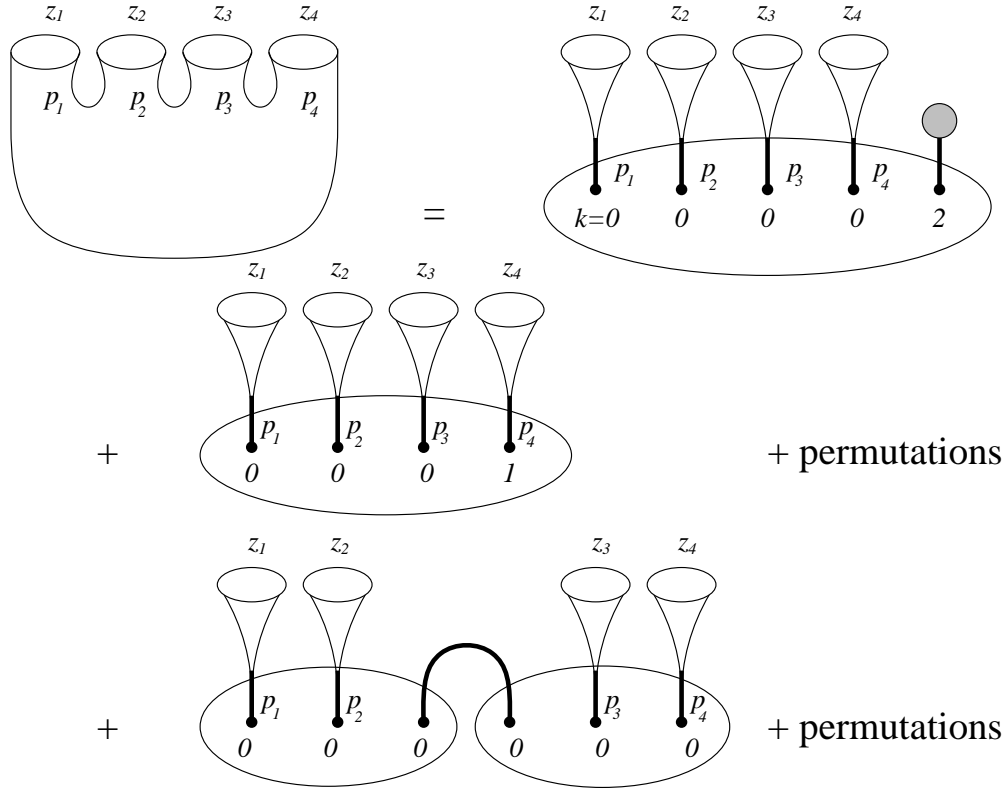
\ifig\fourloopgenuszero{Four-loop genus-zero amplitude}{4loopgenus0}

In the limit $\beta \to 1$ of one-dimensional target space the fusion
coefficient represents a
  periodic delta-function, $ C^0_{p_1\cdots p_n}= \delta ^{(2)}(
p_1+\cdots+p_n)$,
and  eq. \ivan \  reproduces the four-loop amplitude found in \ksks
  \eqn\ivmts{\eqalign{ \langle \prod _{s=1}^4    \Psi_{p_s}(z_s)\rangle_0=
 \delta^{(2)}\Big(\sum_{i=1}^4 p_i \Big) & \Bigg[ 4- 2|p_1+p_2|
-2|p_1+p_3|-|p_1+p_4|
\cr &  + \sum_{s=1}^4  \Big( p_s^2 - {\p^2\over\p\tau_s^2}
   \Big)   \Bigg]
\prod_{s=1}^4   {\sinh (1-p_s)\tau_s \over\sin \pi p_s \sinh \tau_s}
 .\cr}}

\newsec{Conclusions}
 In this work, we have constructed  a string field diagram technique  for
$C\le 1$ backgrounds.
An essential feature of this diagram technique is the factorization of the
  vertices  into a matter-dependent
(fusion coefficient) and
gravity-dependent (intersection number) parts.
We have found a minimal representation of the  Feynman rules characterized by
discrete quantum numbers, which  allows to efficiently calculate the
loop  amplitudes on a surface with arbitrary genus.

The  discreteness of the target space was crucial for the derivation of our
Feynman rules.
However, as it was argued in \ksks, there might be a one-to-one correspondence
between the string theories
with discrete and continuous target spaces. Such a  correspondence would
confirm the idea of Klebanov and Susskind about the appearance of a minimal
length in the target space \ref\KS{I. Klebanov and L. Susskind, \np B 309
(1988) 175}.
In ref. \ksks\ the   target spaces $A_{\infty} \sim \Z $ and $\R$ were
compared. We expect that, more generally,   every string  theory with discrete
target space can be mapped onto a string theory with   one-dimensional space of
 orbifold type.
However, after translating the Feynman rules into the continuum language, the
vertices will loose their  beautiful factorized form and  the so called
"special states" with integer momenta will appear as a remnant of the
periodicity in the momentum space of the original vertices and propagator.

Another important issue to be addressed is to  derive the string field
perturbative expansion  from
 more fundamental  geometrical principles,
without reference to a matrix model. For this one has to find the underlying
algebraic structure and the related symmetry.
 One of the possible approaches relies on the    $W$-algebra symmetry \dvv .
However, it seems that a more natural  structure, following directly from the
loop equations, is described by a
 direct sum of Virasoro symmetries,
one for each point of the target space
\ref\shiba{N. Ishibashi and H. Kawai, \pl B 322 (1994) 67}  \Iope .

Finally, let us mention that our discrete Feynman rules can be readily extended
to the open string sector  using the explicit expressions for the interactions
of open strings derived in ref. \Iope.

   \bigbreak\bigskip\bigskip\centerline{{\bf Acknowledgments}}\nobreak
We thank Bertrand Eynard, Philippe Di Francesco, Charlotte Kristjansen, Volodya
Kazakov and Matthias Staudacher for valuable discussions.
 \listrefs

  \bye